\documentclass[a4paper,11pt]{article}
\pdfoutput=1 
\usepackage{jcappub} 
\usepackage{amsfonts,graphics,color}
\usepackage{graphicx}
\usepackage{amsmath}
\usepackage[T1]{fontenc} 
\usepackage{diagbox}
\usepackage{float}

\newcommand{\black}{\color{black}}

\title{
Can a gamma-ray dim radio blazar produce a 200-PeV neutrino? \\
The case of PMN~J0606$-$0724 and KM3-230213A}

\author[a,b,1]{Polina Kivokurtseva\note{Corresponding author.}}
\author[a,b]{and Sergey Troitsky}

\affiliation[a]{Institute for Nuclear Research of the Russian Academy of Sciences,\\ 60th October Anniversary Prospect 7a, Moscow 117312, Russia}
\affiliation[b]{Faculty of Physics, M.V. Lomonosov Moscow State University,\\ 1-2 Leninskie Gory,  Moscow 119991, Russia}

\emailAdd{kivokurtceva.pi19@physics.msu.ru}
\emailAdd{st@ms2.inr.ac.ru}

\abstract{An extremely energetic muon has been recently detected by the Cubic Kilometre Neutrino Telescope (KM3NeT), indicating the observation of a neutrino with the estimated energy of $\left( 2.2^{+5.7}_{-1.0} \right)\times 10^{17}$~eV. Radio blazar PMN~J0606$-$0724, not detected in gamma rays, is located within the reported error region of the neutrino arrival direction, and was flaring at the time of the event. Here we demonstrate that the neutrino could be produced in a photohadronic interaction in its  millimeter  radio core, a parsec-scale region at the first collimation point of the jet. The necessary proton power is of order of the source's photon luminosity, and protons can be accelerated to the required energies in the core, while high-energy gamma rays cannot leave the source because of intense production of electron-positron pairs. Expected contribution of the population of similar flaring sources matches non-observation of  other energetic events by neutrino telescopes. }


\usepackage{amsfonts,graphics,color}
\usepackage{graphicx}
\usepackage{color}
\usepackage{hyperref}

\begin{document}
\maketitle

\flushbottom


\section{Introduction}
\label{sec:intro}
Considerable attention has been attracted by the detection \cite{KM3NeT:event} of an unusually energetic event KM3-230213A by the KM3NeT neutrino telescope in the Mediterranean Sea during the experiment's deployment stage. With its estimated energy of $\left( 1.2^{+1.1}_{-0.6} \right)\times 10^{17}$~eV, the detected muon was the highest-energy elementary particle ever directly registered. At the confidence level better than 99.99\%, it was produced in the interaction of an extraterrestrial neutrino in the surrounding sea or rock. The estimated energy of this neutrino is then $\left( 2.2^{+5.7}_{-1.0} \right)\times 10^{17}$~eV, while its astrophysical origin remains unknown. A persistent source of neutrinos of that high energy is difficult to reconcile with null results of experiments with larger exposures, namely the Pierre Auger Observatory \cite{Auger:limit2022}, IceCube \cite{IceCube:VHE}, and Baikal Gigaton Volume Detector (GVD) \cite{Baikal:VHE-limits}; see also \cite{KM3NeT:agnostic,titans}.

Among potential sources of the most energetic particles in the Universe are blazars, defined as active galactic nuclei with relativistic jets observed at very small angles. The most direct method to select blazars observationally is to obtain their images at the ultimately achievable angular resolutions with very long baseline radio interferometry (VLBI). A fraction of the VLBI-selected blazars manifest themselves also in high-energy gamma radiation. 

It was expected that the neutrino flux from blazars peaks at energies well above PeV, see e.g.\ Refs.~\cite{BerezinskyNeutrino77,Mannheim1992-FSRQ,Stecker:1991vm,AtoyanDermer,NeronovWhich,P1} and more recent Refs.~\cite{P2,P3,P4}.

During the last decade, evidence  has been mounting suggesting  that blazars contribute to the observed TeV--PeV neutrino flux \cite{IceCube-0506he,Kadler-blazar,neutradio1,neutradio2023}; for discussion and more references see e.g.\ Refs.~\cite{ST-UFN1,ST-UFN2}.  The observational situation regarding the contribution of blazars to the astrophysical neutrino flux remains inconclusive, and inferred level of the contribution varies with the definitions of blazar sample, e.g.\ \cite{IceCube:MeVblazars}, and with the treatment of systematic uncertainties \cite{IceCube:radio-blazars}.  

In Ref.~\cite{KM3NeT:blazars}, the KM3NeT collaboration, jointly with several groups of astronomers,  searched  for potential blazar counterparts of KM3-230213A. One of the sources within the error region of the neutrino arrival direction, PMN~J0606$-$0724, was in the flaring state at the moment of the neutrino arrival, as indicated from radio flux density light curves obtained within blazar monitoring programs by the Owens Valley Radio Observatory (OVRO) 40-m Telescope \cite{OVROmonitoring} and by the 600-m Radio Astronomical Telescope of the Academy of Science (RATAN) in Russia \cite{RATANmonitoring1,RATANmonitoring2}. The chance probability of a random spatiotemporal coincidence of the neutrino event with the flare was estimated as $2.6 \times 10^{-3}$ \cite{KM3NeT:blazars}.

The purpose of the present work is to study in more detail, whether this particular source, never detected in gamma rays, may be  capable of producing extremely energetic neutrinos. In Sec.~\ref{sec:data}, we collect and discuss available multiwavelength observational data on PMN~J0606$-$0724, which are used to estimate parameters of the emitting region in the rest of the paper. Section~\ref{sec:neutrino} discusses the proposed mechanism of the neutrino production. In Sec.~\ref{sec:neutrino:mechanism}, we frame the previously proposed \cite{neutradio2,cores} mechanism of TeV to PeV  neutrino production in stationary features in blazar jets, so-called millimeter radio cores  \cite{Daly_Marscher_1988}.  In Sec.~\ref{sec:neutrino:estimates}, we  present  order-of-magnitude estimates demonstrating that the  MC  scenario can work in PMN~J0606$-$0724 despite the unusual energy of the neutrino  and provide for the implied neutrino flux level. Sec.~\ref{sec:neutrino:numerical} presents the results of a  detailed quantitative analysis of the  same  scenario in the context of a  numerical leptohadronic model of the spectral energy distribution (SED) of this source, confirming our estimates. Making use of constraints from diffuse astrophysical neutrino fluxes, we characterize, in Sec.~\ref{sec:population}, the population of similar extremely energetic blazars, contributing to the full-sky flux. We briefly summarize and discuss our results in Sec.~\ref{sec:concl}.

\section{PMN~J0606$-$0724: observational data}
\label{sec:data}
\subsection{General information and multiwavelength fluxes}
\label{sec:data:MWL}
Being a bright radio source, PMN~J0606$-$0724 was identified as a blazar and observed multiple times at various wavelengths. It has been selected as a potential high-energy source in the Candidate Gamma-Ray Blazar Survey (CGRaBS) \cite{CGRaBS}, but has never been detected in gamma rays. Its redshift, $z=1.277$, was determined in dedicated spectroscopical observations \cite{CGRaBS}. For the commonly accepted cosmological parameters, $h=0.7$ and $\Omega_\Lambda=1-\Omega_{\rm M}=0.685$ \cite{PDG}, the luminosity distance to the source is $d_{\rm L}\approx 8.8$~Gpc. 

To estimate steady-state fluxes across different bands, we made use of the Markarian Multiwavelength Data Center (MMDC) \cite{MMDC} data collection tool and obtained flux measurements from GLEAM\footnote{See the cited papers for explanation of acronyms.} \cite{GLEAM}, TGSS \cite{TGSS}, RACS \cite{RACS}, NVSS \cite{NVSS}, VLASS \cite{VLASS}, PMN survey \cite{Parkes-MIT-NRAO_PMN}, CRATES \cite{CRATES}, Australia Telescope \cite{Australia_Telescope}, ALMA \cite{ALMA}, WISE \cite{WISE}, NEOWISE \cite{NEOWISE}, PanSTARRS \cite{Pan-STARRS}, Gaia \cite{Gaia}, ZTF \cite{ZTF}, ASAS-SN \cite{ASAS-SN}, and eROSITA \cite{eROSITA}, together with data provided by SPECFIND \cite{SPECFIND}. We supplement the data provided by MMDC with those presented in Ref.~\cite{KM3NeT:blazars}, which include data collected from ATLAS \cite{ATLAS} and CRTS \cite{CRTS} observations, newer data from ALMA \cite{ALMA}, dedicated analyses of data from OVRO and RATAN-600, and upper limits from Fermi LAT.

The direction to the blazar is moderately close to the Galactic plane (the Galactic latitude $b\approx 11^\circ$) and passes through a system of molecular clouds \cite{KM3NeT:Galactic}. Not surprisingly, the Galactic extinction is important in this direction. Comparison of various extinction models \cite{GALExtin} demonstrates a significant scatter in the V band correction, from 0.179$^{\rm m}$ \cite{extinction-Amores} to 2.12$^{\rm m}$ \cite{Schlafly2011}. We use the H~I column density from Ref.~\cite{HI-column} and convert it to the V-band extinction $A_{\rm V}$ with the relation of Ref.~\cite{HI-to-extinction} to get $A_{\rm V}=0.743^{\rm m}$; the color correction of Ref.~\cite{Schlafly2011} is used to recalculate the extinction for other optical and infrared bands. 

\begin{figure}
\centerline{\includegraphics[width=0.7\linewidth]{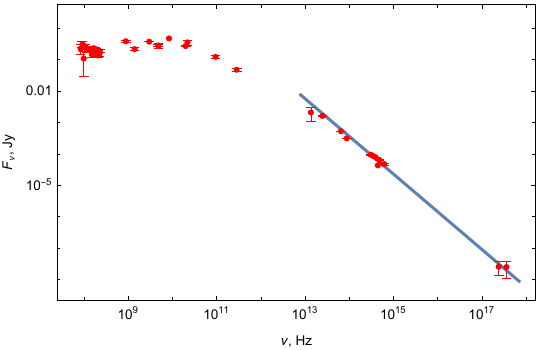}}
\caption{\label{fig:SED1} 
Time-averaged off-flare radio to X-ray SED of PMN~J0606$-$0724. The blue line shows the power-law approximation (\ref{eq:points_flux}) used in Sec.~\ref{sec:neutrino:estimates}.}
\end{figure}

The broadband spectral energy distribution (SED), averaged over observations outside of the period of the neutrino-associated flare, Sec.~\ref{sec:data:flare}, is presented in Fig.~\ref{fig:SED1}. 
We analyze the SED in more detail below in Sec.~\ref{sec:neutrino:numerical}, and use the fluxes in Sec.~\ref{sec:neutrino:estimates} for our estimates.

\subsection{Neutrino-associated flare}
\label{sec:data:flare}
The record-breaking event KM3-230213A \cite{KM3NeT:event} was detected on MJD~59988 and interpreted as an astrophysical neutrino with energy $7.2\times 10^{16} \lesssim E_\nu/\mbox{eV} \lesssim 2.6 \times 10^{18}$ (90\% CL). Large and asymmetric uncertainties in $E_\nu$ are typical for energetic muon-track events, see e.g.\ Ref.~\cite{ST-UFN1}. The origin of the neutrino is uncertain: it can either be associated with one of point sources within the large error circle of the arrival direction \cite{KM3NeT:blazars,palmisano2025exploringultrahighenergyneutrino}, or come from a purely diffuse component produced by interactions of energetic cosmic rays with ambient matter or radiation \cite{KM3NeT:Galactic,KM3NeT:cosmogenic}. Thanks to long-term radio monitoring of large homogeneous samples of blazars by OVRO and RATAN-600, Ref.~\cite{KM3NeT:blazars} identified a strong radio flare in one of the blazars potentially associated with KM3-230213A, namely in PMN~J0606$-$0724, and estimated the probability of a chance coincidence of the event with the flare as $0.26\%$. Other potential sources in this direction in the sky, for which such pronounced flares were not observed, have been discussed in Refs.~\cite{Dzhatdoev:KM3_230213A,Neronov:KM3_230213A,yuan2025accretionflareinterpretationuhe}. We stress that regular observations are performed only for a fraction of blazars, and even more sources could become potential candidates if they were monitored. 

During the time interval $\Delta t$ of the flare, one event with the energy in the 90\%~CL interval $\Delta E_\nu$, quoted above, was observed by KM3NeT. Presumably\footnote{We assume that corresponding events, if any, would have been reported in Refs.~\cite{IceCube:VHE,Baikal:VHE-limits}.}, no events with these energies were observed by other neutrino telescopes, IceCube and Baikal-GVD. One can thus estimate the corresponding energy flux as
$$
\mathcal{F}_\nu = \frac{E_\nu^2\, N}{\Delta E_\nu\, \Delta t\, A},
$$
where $N$ is the expected Poisson mean of the number of detected events and $A=A_{\rm KM3}+A_{\rm IC}+A_{\rm GVD}$ is the total effective area of KM3NeT ($A_{\rm KM3}$), IceCube ($A_{\rm IC}$) and Baikal-GVD ($A_{\rm GVD}$) at the best-fit neutrino energy $E_\nu$. For one observed event and zero expected background, $N=1^{+1.35}_{-0.62}$ (68\%~CL), see e.g.\ Ref.~\cite{PDG}. Following Ref.~\cite{Neronov:KM3_230213A}, we take $A_{\rm KM3}=400~{\rm m}^2$ and $A_{\rm IC}=6000~{\rm m}^2$. Baikal-GVD has the instrumented volume of $\sim 0.7$~km$^3$, but published reports discuss only the cascade channel, for which the effective area is much smaller than for muon tracks. Ref.~\cite{Baikal:VHE-limits} reports the effective area $\approx 1300$~m$^2$ at 220~PeV for a 12-cluster configuration, which we scale to match the 10-cluster instrument operating in February 2023 to get $A_{\rm GVD}=1080~{\rm m}^2$. In this way, we obtain
\begin{equation}
    \mathcal{F}_\nu = \left(\frac{\Delta t}{\mbox{yr}} \right)^{-1} 
    \left(1.3^{+1.8}_{-0.8} \right) \times 10^{-11}~\frac{\mbox{erg}}{\mbox{cm}^2 \, \mbox{s}}.
    \label{eq:Fnu}
\end{equation}

For the purposes of the present work, we estimate the neutrino flux from PMN~J0606$-$0724 from a single observation by the Global Neutrino Network \cite{Spiering:GNN}, combining exposures of its three experiments. This exposure is however dominated not by KM3NeT, and this may be considered as a moderate statistical tension. As we discuss below in Sec.~\ref{sec:population}, there exists a population of similar blazars, contributing to the diffuse neutrino flux. Since blazars are located at cosmological distances, their distribution in the sky is isotropic, and the estimate of the probability that the first event was detected by KM3NeT from Ref.~\cite{KM3NeT:event}, $\sim 2\%$, is applicable.

\begin{figure}
\centerline{\includegraphics[width=0.8\linewidth]{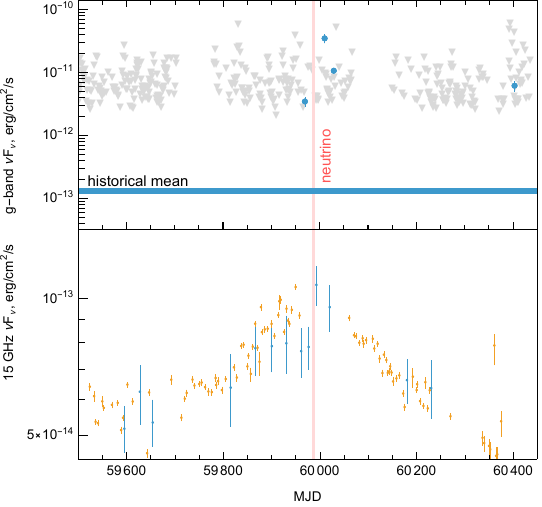}}
\caption{\label{fig:lightcurves} Optical and radio light curves of PMN~J0606$-$0724. Top panel: g-band, ASAS-SN detections  (blue) and $5\sigma$ upper limits (gray) together with the historical PanSTARRS mean and standard deviation (line thickness). Bottom panel: 15~GHz, OVRO (yellow) and RATAN-600 (blue) fluxes. The day of the neutrino event is marked with the pink vertical line.}
\end{figure}

To estimate the electromagnetic fluxes during the flare, we need to select the corresponding time interval, which is ambiguous. In Sec.~\ref{sec:neutrino}, we will keep the effective flare duration as a free parameter. We choose the period of $\pm 45$ days around the event as the approximate period of the flare at half-maximum, cf.\ Fig.~\ref{fig:lightcurves}, and select data with observation epochs in this interval. 

Lacking a neutrino alert from KM3NeT, no target-of-opportunity observations were performed, and the data are scarce for this period. Ref.~\cite{KM3NeT:blazars} presents OVRO, RATAN-600, ALMA and WISE data as well as Fermi-LAT upper limits. In addition, we found that PMN~J0606$-$0724 was detected by ASAS-SN during the flare, though its flux is normally well below this instrument's sensitivity. In Fig.~\ref{fig:lightcurves},
we present the light curve of the radio flare and simultaneous ASAS-SN observations, compared to the historical mean flux from PanSTARRS observations at the same wavelength.

The information on the neutrino-associated flare of PMN~J0606$-$0724 is dominated by the radio monitoring data. The neutrino arrival time, unluckily, coincided with a technical failure of the OVRO 40-meter telescope, which caused the largest, 88-day, gap in the 18-year blazar monitoring program (A.C.~Readhead, private communication). However, OVRO data clearly trace the rise and the fall of the flare, while the peak was caught by the RATAN-600 monitoring observations.

\section{Neutrino production}
\label{sec:neutrino}
\subsection{Mechanism}
\label{sec:neutrino:mechanism}
From early days of neutrino astronomy to the publication of first IceCube results, models of neutrino production in active galactic nuclei were mostly based on photohadronic interactions in the vicinity of supermassive black holes, see e.g.\ Refs.~\cite{BerezinskyNeutrino77,BerezGinzb,Stecker:1991vm,AtoyanDermer,KalashevAGN}, and more references in \cite{ST-UFN1}. Because of the thermal spectra of target photons from the accretion disk, the predicted neutrino spectra also spanned a limited range of energies, mostly in the PeV band. This scheme was challenged by subsequent IceCube data, which indicated that blazars can produce neutrinos with energies from TeV to PeV bands, in some cases even in one and the same source, see e.g.\ Refs.~\cite{IceCube:TXS-Science2,neutradio2,IceCube:10yr-map,Suray-flares}.  Further studies, e.g.\ \cite{FedNat}, concluded that it is hard to explain the production of neutrinos in the radiative blazar zone, and suggested that multi-zone models may be required.  A possible solution was to consider neutrino production right in the developed  relativistic jet, where nonthermal radiation of the jet itself \black provides for the target-photon spectrum spanning many decades in energy \cite{neutradio2}. In particular, Ref.~\cite{cores} suggested a standing shock close to the jet base, known as the millimeter radio core  (hereafter,  MC ) \cite{Daly_Marscher_1988}, as a possible site of interactions of relativistic protons with target photons, producing neutrinos of various energies. While direct observations of these shocks are challenging\footnote{See however Refs.~\cite{EHT-3C279,EHT-J1924-2914}.} because of synchrotron self-absorption and the lack of millimeter-wavelength VLBI observations, modeling of jet properties often favors the existence of such stationary or slow features in the central parsecs of blazars, see e.g.\ Ref.~\cite{Egor}. This mechanism is discussed in detail in Ref.~\cite{cores}. In what follows, we apply it to the particular case of PMN~J0606$-$0724.

It is important to recall that the MC  is different from the discrete blobs propagating down the jet, which were studied e.g.\ in Refs.~\cite{P1,P2,P3,P4}. The  MC  is not a physical clump of matter, but rather a stationary, geometrical region through which the relativistic plasma flows. As it was discussed in previous works, e.g.\ \cite{Daly_Marscher_1988,cores}, in detail, the Lorentz factor (and hence the Doppler factor, assuming no jet bending) is slightly lower for the plasma in the  MC  compared to that of blobs moving in the fully developed jet. Still, the motion is ultrarelativistic, with typical expected Doppler factors of $2 \lesssim \delta_c \lesssim 5$, see Ref.~\cite{cores} and references therein. With a few known exceptions, blazar  MC  dominates electromagnetic radiation of the jets. The  MC  is located at the parsec-scale distance from the central black hole and has itself the size of order $\sim 1$~pc. 

Throughout the rest of the paper, we use standard notations in which $\mathcal{F}$ [erg/(cm$^2\,$s)] denotes the energy flux, $L$ [erg/s] denotes luminosity or power; $E_\nu$, $E_p$, $E_\gamma$ denote energies of neutrino, proton, and photon, respectively. Quantities in the production-region frame, which are boosted with the relativistic Doppler factor of the  MC , $\delta_c$, and blueshifted by the cosmological factor $(1+z)$, are primed, while those in the blazar central machine frame, that is only blueshifted, are starred:
$$
E'=\frac{1+z}{\delta_c}E, ~~~ E^\star = (1+z)E,
$$
where $E$ is the energy in the observer frame. Target photons are assumed to be isotropic in the production-region frame.

\subsection{Estimates}
\label{sec:neutrino:estimates}

In this subsection, we give order-of magnitude estimates demonstrating that, despite the unusually high energy of the KM3~230213A neutrino, the mechanism of neutrino production in the MC  of a blazar jet  -- developed for TeV to PeV neutrinos -- is still expected to work.Here, we present analytical estimates obtained under various symplifying assumptions, described in the remaining part of Sec.~\ref{sec:neutrino:estimates}.  The reader is referred to Sec.~\ref{sec:neutrino:numerical} for a more realistic numerical analysis.

The photohadronic  mechanism for neutrino production requires two ingredients, namely, accelerated protons and target photons. 
Exclusively  for the estimates in this subsection, we recall that the $p\gamma$ cross section peaks strongly at the energy corresponding to the resonant creation of a $\Delta^+$ baryon in the intermediate state. For this case, energies of the target photon, $E'_\gamma$, and of the projectile proton, $E'_p$, are related by $E'_\gamma E'_p\simeq m^2_\Delta$, where $m_\Delta\approx 1.23$~GeV is the $\Delta$-particle mass. 
In addition, the average energy of the produced neutrino is $E'_\nu \approx E'_p/20$, see e.g.\ \cite{ST-UFN1} and references therein. 
\subsubsection{Maximal energy of protons}
\label{sec:neutrino:estimates:Eproton}
The required proton energy is thus
$$
E'_p \simeq 20 \,E_\nu \frac{1+z}{\delta_c}\approx \frac{1}{\delta_c}\left( 1.0^{+2.6}_{-0.5} \right)\times 10^{19}~\mbox{eV}.
$$
Cosmic-ray protons with these high energies are ubiquitous, see e.g.\ Ref.~\cite{Anchordoqui:CR-review}, but it is presently unclear where they are accelerated. For our purposes, we consider two scenarios.

\textit{1. Acceleration in the jet,} either in the mm  MC  itself, or in the faster part before collimation. Basic requirements for the physical conditions in the acceleration region have been reviewed e.g.\ in \cite{PtitsynaHillas,Simon:constraints}. In particular, the geometric, or Hillas, criterion reads
$$
    E'_p\lesssim 9.3 \times 10^{19}~\text{eV}~\left(\frac{B'}{0.1\black~\rm G}\right) \left(\frac{R'}{\rm pc}\right),
$$
where $R'$ is the size of the acceleration region, and $B'$ is the magnetic field there. The constraint from diffusive synchrotron losses requires 
$$
    E'_p\lesssim 2.9 \times 10^{21}~\text{eV}~\left(\frac{B'}{0.1\black~\rm G}\right)^{-2} \left(\frac{R'}{\rm pc}\right)^{-1}.
$$
The value of $R'$ may be estimated from the variability timescale $\Delta t_{\rm var} \sim 1$~yr,
$$
R' \sim \delta_c \, \frac{\Delta t_{\rm var}}{1+z} \sim  0.13 \black \, \delta_c \,\frac{\Delta t_{\rm var}}{\rm yr}  ~{\rm pc}. 
$$
This size fits nicely the typical scale of the jet base or the MC. Taking $R'\sim 0.2$~pc as a benchmark, we arrive at the condition
\begin{equation}
 \frac{0.5}{\delta_c} \lesssim \frac{B'}{0.1\black~\rm G} \lesssim  38 \black\sqrt{\delta_c}.   
 \label{eq:B-accel-in-jet}
\end{equation}
This condition can easily be satisfied for reasonable values of the magnetic field in the  MC \black. 

Interaction losses are a bit more tricky, since one needs a balance between proton acceleration and neutrino-producing interactions. Calculations presented in Ref.~\cite{Simon:constraints}, see their Fig.~1, demonstrate that the losses marginally allow the acceleration of protons to the required energies; we will return to this point in the next subsection.

\textit{2. Acceleration before the jet launching,} in a vacuum gap of the magnetosphere of the central black hole. Here, electric field may accelerate particles in one shot, which helps to avoid additional energy losses. This setup has been explored in detail in Ref.~\cite{PtitsynaNeronov}. Following this reference, we estimate the maximal energy of protons, which now refers to $E^\star_p=E'_p \delta_c$. To this end, we use their Eqs.~(19)--(21) to estimate the vacuum gap size, assuming the typical value of the magnetic field \cite{PtitsynaHillas} on the black-hole horizon, $B^\star \sim 10^4~{\rm G}~M_8^{-1/2}$, for the black-hole mass $M_{\rm BH} \equiv M_8 \,10^8 M_\odot$, and taking into account the synchrotron peak energy for the ambient photons, $\epsilon_{\rm IR}\sim 0.01$~eV. The maximal energy is then
$$
E_p^\star \sim 1.3 \times 10^{13}~\mbox{eV}\, M_8^{1/2},
$$
so that the requirement $E_p^\star\approx 10^{19}$~eV pushes $M_{\rm BH}$ to uncomfortably high values $\sim 10^{12} M_\odot$.

We conclude that acceleration of protons to the energies required for producing a 220-PeV neutrino is possible in the jet of PMN~J0606-0724, but not in the magnetosphere vacuum gap of its black hole. 

\subsubsection{Target photons}
\label{sec:neutrino:estimates:photons}
Following Ref.~\cite{cores}, we consider a cylindrical  MC  region extended along the jet direction with typical values of the length $l' \approx 1.2$~pc and radius $r' \approx 0.2$~pc. 

The energy-dependent specific number density of photons, $n'_\gamma(E'_\gamma)$  (recall that primed quantities are in the MC frame, which moves relativistically with the Doppler factor $\delta_c$)\black, measured in cm$^{\text{-}3}$eV$^{\text{-}1}$, is related to the energy flux $\mathcal{F}(E_\gamma)$, observed at the Earth, measured in erg$\cdot$cm$^{\text{-}2}$s$^{\text{-}1}$, as  \cite{Dermer} 
\begin{equation}
E'_\gamma n'_\gamma(E'_\gamma)=
\frac{4d_L^2\mathcal{F}}{\delta_c^4 E'_\gamma r'^2}
\left( \frac{r'}{l'}\right)^{2/3}.
\label{eq:F-to-n}
\end{equation}
Based on the observations of PMN~J0606$-$0724, cf.\ Fig.~\ref{fig:SED1}, we approximate $\mathcal{F}(E_\gamma)$ for the energies of interest for target photons as
\begin{equation}
\mathcal{F}\left( E_\gamma \right)= \mathcal{F}_0 \left(\frac{E_\gamma}{E_0}\right)^{-0.2},
\label{eq:points_flux}    
\end{equation}
where $\mathcal{F}_0=2.9 \times 10^{-13}$~erg$\cdot$cm$^{\text{-}2}$s$^{\text{-}1}$ and $E_0=0.41 $~eV. 

Neutrinos are produced in $p\gamma$ interactions, whose cross section peaks at the $\Delta$-baryon resonance. Neglecting off-resonance contributions, the required proton energy in the emitting-region frame is $E'_{p} \approx 20E_{\nu} (1 + z)/\delta_c$. This also fixes the most favorable   energy of the target gamma rays as $E'_\gamma \approx m_\Delta^2/E'_{p}$. For the observed energy of these target photons, one obtains
$$
E_\gamma \simeq \frac{\delta_c^2}{(1+z)^2} \frac{m_\Delta^2}{20 E_\nu} \equiv \delta_c^2 E_{\gamma0},
$$
where $E_{\gamma0}= 0.066^{+0.17}_{-0.033}$~eV for the KM3~230213A neutrino energy and the PMN~J0606$-$0724 redshift. This typical observed energy of target photons is covered by observations, and we use the approximation Eq.~(\ref{eq:points_flux}) to transform Eq.~(\ref{eq:F-to-n}) into
$$
E'_\gamma n'_\gamma(E'_\gamma)=
\frac{4d_L^2\mathcal{F}_0}{E_0 (1+z)\delta_c^{5.4} E'_\gamma r'^2}
\left( \frac{r'}{l'}\right)^{2/3}
\left( \frac{E_{\gamma 0}}{E_0}\right)^{-1.2}
$$
$$
\approx \left( 5.6^{+16}_{-3.2}\black \right) \times 10^{-13}\, \mbox{cm}^{-3}
\, \delta_c^{-5.4} \left( \frac{r'}{\mbox{pc}}\right)^{-2} \left( \frac{6r'}{l'} \right)^{2/3}.
$$

\subsubsection{Multimessenger luminosities}
\label{sec:neutrino:estimates:lumi}
In the $\Delta$-resonance process, a proton with energy $E'_p$ gives either another proton and a neutral pion $(p\pi^0)$, or a neutron and a charged pion $(n\pi^+)$. Isospin conservation requires that the branching ratios of the two channels are related by
$$
\mbox{Br}\!\left(p\pi^0\right) = 2\, \mbox{Br}\!\left(n\pi^+\right) .
$$
Therefore, 1/3 of the initial energy goes to the charged channel, producing three neutrinos, each with the energy $\sim E'_p/20$. The total proton luminosity  needed to account for the neutrino emission can thus be estimated as  $L'_p \sim 20 L'_\nu/\xi$, where $\xi$ is the fraction of protons which interact. Note that luminosities, measured in erg/s, are relativistic invariants, so we do not distinguish $L'$ and $L^\star$.

The fraction $\xi$ is related to the proton optical depth,
\begin{equation}
    \tau_{p\gamma} =l'_1 \sigma E'_\gamma n'_\gamma,
    \label{eq:tau-pgamma}
\end{equation}
as
$
\xi = 1 - \exp(-\tau)
$. 
Here, $\sigma \approx 500~\mu$b is the $p\gamma$ cross section at the $\Delta$ resonance, and for the values of parameters estimated above in Sec.~\ref{sec:neutrino:estimates:Eproton}, \ref{sec:neutrino:estimates:gamma}, Eq.~(\ref{eq:tau-pgamma}) transforms to $\tau \approx 260/\delta_c^{5.4}$. The most efficient transfer of the proton power to neutrinos occurs at $\tau \sim 1$, that is $\delta_c \approx 2.8$. We note that the values $2 \lesssim \delta_c \lesssim 5$ are expected for the  MC \cite{cores}.  For the remaining part of Sec.~\ref{sec:neutrino:estimates}, we use $\delta_c=2.8$ as a benchmark value for our estimates.

Making use of the arguments similar to the estimate of the neutrino flux, Eq.~(\ref{eq:Fnu}), one obtains the neutrino luminosity, 
$$
L_\nu = 
\frac{1}{\delta_c^4}\, 4\pi\,d_L^2 \frac{E_\nu}{A\Delta t}
=
\left(\frac{2.8}{\delta_c} \right)^4 
\left(\frac{\mbox{yr}}{\Delta t} \right)
\left( 2.3^{+5.9}_{-1.0}\right)
\times 10^{46} 
~
\frac{\mbox{erg}}{\rm s}.
$$
The proton power is estimated as
\begin{equation}
L_p \sim \black 
20L_\nu/\xi\black=\frac{1}{\xi}\black
\left(\frac{2.8}{\delta_c} \right)^4 
\left(\frac{\mbox{yr}}{\Delta t} \right)
\left( 4.5^{+11.8}_{-2.1}\right)
\times 10^{47} 
~
\frac{\mbox{erg}}{\rm s}.
\label{Eq:Lp-estimate}
\end{equation}
These values are of the same order as the Eddington luminosity of a supermassive black hole typical for bright quasars,
$$
L_{\rm Edd} = 
1.3
\times 10^{47} \frac{M_{\rm BH}}{10^9 M_\odot}
~
\frac{\mbox{erg}}{\rm s}.
$$
The exact value of $M_{\rm BH}$ for PMN~J0606$-$0724 is not known; however, unless $M_{\rm BH} \gtrsim 10^9 M_\odot$ or one of the assumptions of Eq.~(\ref{Eq:Lp-estimate}) is relaxed, the estimated proton luminosity exceeds $L_{\rm Edd}$.

\subsubsection{Gamma-ray opacity}
\label{sec:neutrino:estimates:gamma}
Together with neutrinos, $p\gamma$ interactions produce a number of energetic gamma rays coming from decays of neutral pions and having energies two times higher than neutrinos, that is for the event of interest about
$$
E'_{\gamma1} = 
\left(\frac{2.8}{\delta_c} \right)
\left( 3.6^{+10}_{-1.8}\right)
\times 10^{17} 
\,
\mbox{eV}.
$$
We use the index ``$\gamma1$'' to mark energies and luminosities of energetic photons produced in $p\gamma$ interactions, keeping the index ``$\gamma$'' for low-energy target photons. 

 Here, we continue to perform order-of-magnitude estimates and work within the approximations outlined in the beginning of Sec.~\ref{sec:neutrino:estimates}. The gamma rays from $\pi^0$ decays are thus considered essentially monoenergetic. However, e\black{}ven before exit to the extragalactic space, the energetic photons experience intense pair production on low-energy ambient photons in the source, and electrons and positrons further contribute to the development of an electromagnetic cascade. The secondary photons might in principle be detected by Fermi LAT, but the source has not been detected in this band. Following Ref.~\cite{Gorbunov:opt-thick}, let us demonstrate that the non-detection is expected, as the region of $p\gamma$ interaction is optically thick for photons detectable by Fermi LAT, that is having energies $E_\gamma \gtrsim 0.1$~GeV, cf.\ Ref.~\cite{BlandfordLevinson}.

The photons of interest, which may contribute to the putative Fermi-LAT signal either directly, or via intergalactic electromagnetic cascades on the cosmic background radiation, have energies
$$
\frac{1+z}{\delta_c} \cdot 0.1~\mbox{GeV}
\lesssim
E'_{\gamma1}
\lesssim
\frac{2.8}{\delta_c} \cdot 3.6 \times 10^{17}~\mbox{eV}.
$$
In the $\delta$-function approximation \cite{Dermer}, the target photons for the $e^+e^-$ production have energies
$$
\frac{\omega'}{\rm eV}
\sim\frac{5 \times 10^{11}~\rm eV}{E'_{\gamma1}},
$$
which correspond in the observer frame to 
$$
\left(\frac{\delta_c}{2.8}\right)^2 \cdot 4.1 \times 10^{-6}
\lesssim
\frac{\omega}{\rm eV}
\lesssim
\left(\frac{\delta_c}{2.8}\right)^2 \cdot 7.6 \times 10^{3}.
$$
We use again Eq.~(\ref{eq:F-to-n}) to relate the observed flux $\mathcal{F}(\omega)$ to the photon density in the source, $n'_\gamma (\omega')$, and to find that the optical depth $\tau_{\gamma\gamma} = l' n' (\omega') \sigma_{\gamma\gamma}>1$ provided that $\mathcal{F}(\omega) >3.3 \times 10^{-17}$~erg\,cm$^{-2}$\,s$^{-1}$ for photons between decimeter and keV bands. In our estimates here in Sec.~\ref{sec:neutrino:estimates}, we assume that the observed flux of photons with the required energies comes from the MC. In Sec.~\ref{sec:neutrino:numerical}, we present an explicit numerical fit of the entire observed SED with the core emission. \black The latter relation holds, see Fig.~\ref{fig:SED1}, so the density of target photons is sufficient to suppress the flux of gamma rays in the Fermi-LAT band. This is typical for active galactic nuclei in which detectable fluxes of lower-energy neutrinos can be produced by means of the $p\gamma$ mechanism, cf.\ e.g.\ Refs.~\cite{no-gamma-Reimer,no-gamma-Murase}. The lack of association of neutrino and gamma-ray activity of blazars was also found in the data, see e.g.\ Ref.~\cite{ST-UFN1} for a summary of numerous results. In some cases, the gamma-ray flux might even decrease temporarily at the high-energy neutrino arrival time \cite{KunNoGamma}, while the source is flaring in radio \cite{neutradio1,Hovatta-neutrino}. However, one should note that all previous studies dealt with neutrino energies orders of magnitude lower than we discuss here.

One may apply qualitative estimates similar to Sec.~\ref{sec:neutrino:estimates:lumi} to relate the powers emitted in protons and in the energetic secondary gamma rays. In the $\Delta$-resonance approximation, 2/3 of proton interactions produce neutral pions, and each of the latter immediately decays to two photons with energies $E'_{\gamma1}$. This implies
$$
    L_{\gamma1}= \frac{8}{3}L_\nu
    =
\left(\frac{2.8}{\delta_c} \right)^4 
\left(\frac{\mbox{yr}}{\Delta t} \right)
\left( 6.0^{+15.7}_{-2.7}\right)
\times 10^{46} 
~
\frac{\mbox{erg}}{\rm s}.    
$$
At the same time, the steady-state bolometric luminosity from radio to X-ray bands of PMN~J0606$-$0724 may be estimated from the observed SED making use of Eq.~(\ref{eq:F-to-n}),
$$
L_{\rm bol} \approx \left(\frac{2.8}{\delta_c} \right)^4 
1.5
\times 10^{45} 
\,
\frac{\mbox{erg}}{\rm s}.
$$
Even taking into account the $L_{\rm bol}$ increase during the flare, and possible variations in $\Delta t$, these estimates imply $L_{\gamma1} > L_{\rm bol}$. The power $L_{\gamma1}$ is transferred by electromagnetic cascades to the soft gamma-ray band. Like  many  other high-energy neutrino sources, PMN~J0606$-$0724 is expected to be bright in MeV gamma rays, assuming that the neutrino was produced in $p\gamma$ interactions in its MC.

\subsection{Numerical calculation}
\label{sec:neutrino:numerical}

The overall consistency between the estimates made in Sec.~\ref{sec:neutrino:estimates} and the observational data invites a more detailed study, which is performed in this subsection. Here, we calculate numerically  the photon and neutrino spectra, produced by photohadronic interactions in the MC of PMN~J0606$-$0724,  and demonstrate that the observed SED of this blazar  is well described for the source parameters  different but  reasonably close to those estimated in Sec.~\ref{sec:neutrino:estimates}.
We make use of the publicly available simulation code AM$^3$ \cite{AM3}, which self-consistently evolves the spectra of protons, electrons, neutrinos, and photons in a homogeneous, magnetized environment. The code accounts for key radiative and hadronic processes, including synchrotron and inverse Compton emission, photo-hadronic and hadronic interactions, and particle decays. It also tracks energy losses and secondary emissions in a time-resolved manner, making it possible to account for electromagnetic cascades in the source automatically. The code treats all the processes in full, so the delta-function approximation used in Sec.~\ref{sec:neutrino:estimates} is not applicable here. For the calculation of gamma-ray absorption on the extragalactic background light (EBL) during their propagation from PMN~J0606$-$0724 to the Earth, we adopt the Gilmore (2012) fiducial EBL model \cite{Gilmore_2012}. We replace the conventional spherical blob geometry with a cylindrical configuration tailored to our mechanism \cite{cores}. 

As we have pointed out in Sec.~\ref{sec:data:flare}, observational data for the flare period are scarce and do not allow one to reconstruct the SED. Therefore, we use the measurements from the quiescent state, Sec.~\ref{sec:data:MWL}, and scale them by a factor of three, which is an approximate flux enhancement in the only band, radio (cf.\ Fig.~\ref{fig:lightcurves}), reasonably well measured during the flare. This scaling is a simplification because the nature of the flare is not established, and some of the blazar flares change both the amplitude and the shape of SEDs; however, given the lack of data, we use this approximation. We supplement the scaled points with the upper limits on the photon flux measured by \textit{Fermi}-LAT in the energy range $0.1$–$100~\mathrm{GeV}$, as reported in Ref.~\cite{KM3NeT:blazars}, to demonstrate that the model flux remains consistent with the observed limits. We use the 90\% CL limit
$
F_{\rm LAT} < 6.5 \times 10^{-9}~\mathrm{ph\,cm^{-2}\,s^{-1}}
$, which corresponds to 
\[
\mathcal{F}_{\rm LAT} < 9.6 \times 10^{-12}~\mathrm{erg\,cm^{-2}\,s^{-1}}.
\]
for the typical photon energy of 1~GeV.

As it is customary in SED modelling, we treat the injection spectra of electrons and protons (namely, the spectral indices and maximal Lorentz factors) as independent phenomenological parameters. Although the assumption of their equality for different species is sometimes used to reduce the parameter space, electron and proton spectra are not expected to be identical in general because electron radiative cooling and/or species-dependent acceleration mechanism can modify the effective slopes. We verified that imposing the condition of either equal spectral indices or equal maximal Lorentz factors for injected electrons and protons does not yield an acceptable fit for the present data set.

Our one-zone emission region is intended to reproduce the SED above the radio band. At frequencies below $\sim 100$~GHz, the source is expected to be opaque due to synchrotron self-absorption, hence the observed radio emission must arise from larger, downstream jet regions \cite{BlandfordKonigl,MarscherGear}.

\begin{figure}
\centerline{\includegraphics[width=\linewidth]{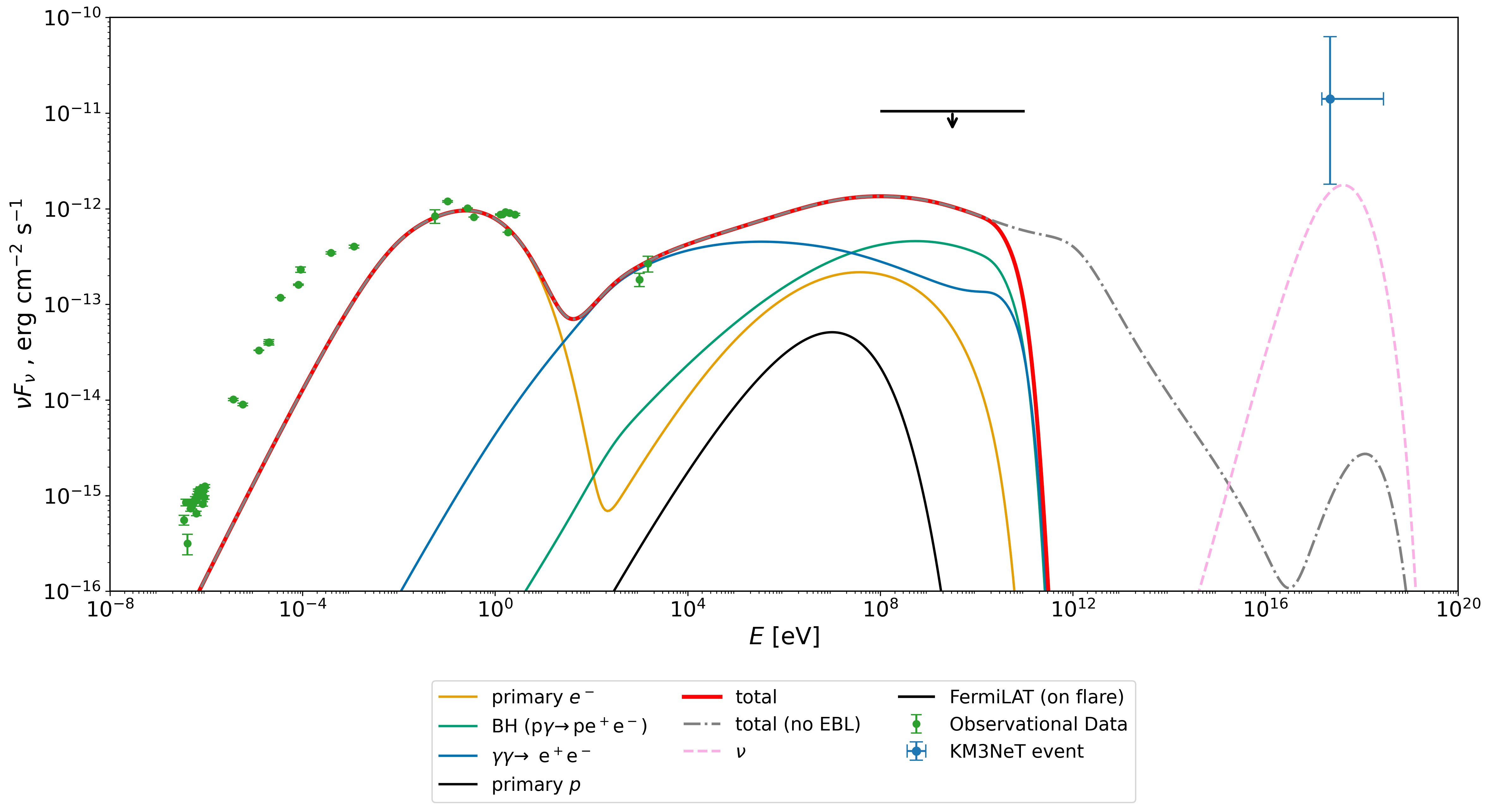}}
\caption{\label{fig:SED} The SED for PMN~J0606$-$0724  during the flare coincident with the neutrino detection.  The total photon spectrum is shown as a thick red curve. The pink dashed line represents the neutrino spectrum (all flavors). The orange line gives the sum of synchrotron and Compton emissions from injected electrons, while the black one is the synchrotron emission from injected protons. The green and blue lines present the synchrotron emission from $e^+e^-$ pairs generated via the Bethe–Heitler process and $\gamma \gamma$ annihilation, respectively. See the text for discussion of the observational data points. The gray dash-dotted line represents the SED without the EBL absorption. The blue cross shows the KM3NeT event flux estimate, assuming $\Delta t = 1 \ \text{yr}$. }
\end{figure}

The SED consistent with the observational data is presented in Fig.~\ref{fig:SED}, and the parameters used for its generation are collected in Table~\ref{tab:parameters}. The KM3NeT event flux shown in Fig.~\ref{fig:SED} is normalized using Eq.~\ref{eq:Fnu} with $\Delta t = 1 \ \text{yr}$. 

\begin{table}[]
    \centering
    \begin{tabular}{ccc}
\hline\hline
Parameter         & \multicolumn{2}{c}{Value} \\
\hline
MC length, $l'$ & \multicolumn{2}{c}{1.5~pc}\\
MC transverse radius, $r'$ & \multicolumn{2}{c}{0.6~pc}\\
magnetic field, $B'$ & \multicolumn{2}{c}{0.05~G}\\
Doppler factor, $\delta_c$ & \multicolumn{2}{c}{5} \\
&Protons: & Electrons: \\
spectral index & 1.0        & 1.8 \\
minimal energy, $E'_{\rm min}$ & $1\times 10^{9}$~eV &  $1\times 10^9$~eV\\
maximal energy, $E'_{\rm max}$ & $2\times10^{18}$~eV         & $10^{10}$~eV\\
total power, $L'$    & $5\times10^{47}$~erg/s & $5\times10^{44}$~erg/s\\
\hline         
    \end{tabular}
    \caption{Parameters used for the SED in Fig.~\ref{fig:SED}.}
    \label{tab:parameters}
\end{table}
The results of this numerical calculations should be treated with care because the synchrotron peak data do not correspond to the flare period but are rescaled from the historical average. We see that the parameters preferred by the SED have reasonable values in the order-of-magnitude agreement with simplified estimates of Sec.~\ref{sec:neutrino:estimates}, including the constraint (\ref{eq:B-accel-in-jet}) for the proton acceleration in the jet. 

In our fit, $L_e$ is fixed by the observed synchrotron SED component, while $L_p$ is determined by the neutrino flux. Large baryon loading is not uncommon in powerful FSRQs. Leptonic SED modelling often implies proton-dominated jet energetics, cf.\ Fig.~2 in Ref.~\cite{CelottiGhisPower} (where $L_p$ is the bulk power in cold protons assuming one proton per emitting lepton).

The choice of the EBL model does not affect the comparison of the modelled SED with observations because the gamma-ray flux is strongly suppressed in the source, before interactions with EBL.

Note that the SED model favors the value of $\delta_c \approx 5$, slightly higher than $2.8$ estimated in Sec.~\ref{sec:neutrino:estimates} as the most efficient for the transfer of proton luminosity to neutrinos. For $\delta_c=5$, Eq.~(\ref{Eq:Lp-estimate}) gives $L_p\approx 4 \times 10^{46}$~erg/s, while the full calculation presented in this subsection results in $5\times 10^{47}$~erg/s. This means that the most optimal regime is not realized in this particular blazar, at the same time illustrating the importance of the full self-consistent, energy-dependent treatment of the particle-physics processes resulting in the neutrino production, as well as of energy losses.

\section{Population of similar sources}
\label{sec:population}
The source we consider, PMN J0606$-$0724, is not special. To estimate the contribution of the population of sources to the full-sky diffuse flux, we assume that there are $N_s$ sources capable to produce neutrinos of these energies, each with the duty cycle $\eta$, defined as the fraction of time during which the source experience a flare sufficiently strong to produce the neutrino flux $\mathcal{F}_\nu$. 
Given the uniform distribution of distant blazars in the sky, one expects the population to contribute to the isotropic diffuse neutrino flux.
Assuming identical sources, one estimates their contribution to the diffuse flux as
\begin{equation}
    \mathcal{F}_{\rm diff} \sim \frac{1}{4\pi} N_s \eta \mathcal{F}_\nu.
\label{eq:pop-flux}
\end{equation}
We do not address here the question of consistency between the observation of one event in KM3NeT and zero events in other experiments at the same energy. This point was extensively discussed e.g.\ in Refs.~\cite{KM3NeT:agnostic,titans,Baikal:VHE-limits}, where the probability to detect the event by the smallest of the telescopes was estimated at the  $\sim 1\%$ level. We follow these studies and use the
combined best-fit steady isotropic diffuse flux based on observation of a single event in KM3NeT, IceCube, Baikal-GVD and Pierre Auger data evaluated in Ref.~\cite{Baikal:VHE-limits} as $7.2 \times 10^{-10}$~GeV$\cdot\,$cm$^{-2}\,$s$^{-1}\,$sr$^{-1}$ per flavor.  Based on the Poisson statistics  \cite{PDG} and multiplying by three flavors, we transform it into  the 90\% CL upper limit of
\begin{equation}
\mathcal{F}_{\rm diff} < 8.4 \times 10^{-9}~\mbox{GeV}\cdot\mbox{cm}^{-2}\,\mbox{s}^{-1}\,\mbox{sr}^{-1}.
\label{eq:diff-flux-limit}
\end{equation}
In the Radio Fundamental Catalog, which contains a complete full-sky sample of radio blasars \cite{RFC}, there are $N_s\sim 600$ sources, which are not detected by Fermi LAT but have a similar or stronger 8-GHz flux compared to PMN J0606$-$0724. 

As mentioned earlier, the non-detection by LAT provides an important constraint: any target photon field dense enough to produce neutrinos would simultaneously and severely attenuate GeV radiation via pair production.

The least studied parameter is the duty cycle, which however may be estimated from the fact that during $\approx 17$~yr of  radio  monitoring  of  the source  by OVRO and RATAN-600, the flare of that amplitude was observed once \cite{KM3NeT:blazars}, so
$$
\eta\sim \frac{\Delta t}{17~\mbox{yr}}.
$$
Note that these values of $\eta$ and $\Delta t$ match those estimated from the radio blazar population studies  based on long-term monitoring of light curves, e.g.\ by Mets\"ahovi, UMRAO \cite{Hovatta:flares} and OVRO \cite{Liodakis:flares} instruments.

Requiring that the total flux (\ref{eq:pop-flux}) does not exceed the diffuse upper limit (\ref{eq:diff-flux-limit}) and using $\mathcal{F}_\nu$ from Eq.~(\ref{eq:Fnu}), we obtain
$$
N_s \lesssim 220^{+310}_{-130},
$$
which demonstrates the overall consistency  (within $1.1\sigma$)  between the single detection from PMN J0606$-$0724 during the flare and non-detection of similar extremely energetic neutrinos from other blazars during the lifetime of all neutrino experiments. 

\section{Conclusions}
\label{sec:concl}
One candidate source of the extremely energetic KM3~230213A neutrino event is PMN~J0606$-$0724, a radio blazar that experienced a strong flare coincident with the neutrino detection. We show that neutrinos at such energies can be produced in this blazar; other potential sources of the event are not considered in our study and therefore are not constrained. Following Ref.~\cite{cores}, we associate the neutrino production site with the millimeter radio core (MC), a parsec-scale, quasi-stationary collimation/recollimation shock in the relativistic jet near its base. Protons can be accelerated to the required energies in the MC region and interact with soft photons, producing energetic neutrinos and gamma rays. The accompanying gamma rays, however, initiate electromagnetic cascades on the same target photon field, and their energy is reprocessed into MeV photons, largely below the Fermi-LAT band. This is consistent with the non-detection of PMN~J0606$-$0724 by Fermi LAT even during the flare. Finally, although this blazar is not unique, the total very-high-energy neutrino emission from the population of similar sources remains below current upper limits on the diffuse neutrino flux at these extreme energies.
\acknowledgments
We are indebted to Yuri Kovalev, Egor Podlesnyi, Anthony Readhead and Grigory Rubtsov for helpful discussions. This work is supported in the framework of the State project ``Science'' by the Ministry of Science and Higher Education of the Russian Federation under the contract 075-15-2024-541. Some of the data used in this work were obtained from MMDC.
\bibliography{blazar}

\end{document}